\input{aipcheck}

\documentclass[
    ,final            
  ]
  {aipproc}

\layoutstyle{6x9}
\usepackage{amssymb}

\begin{document}

\title{{\slshape Fermi}-LAT Detection of Gamma-ray Pulsars above 10 GeV}


\author{P.~M.~Saz Parkinson for the $Fermi$-LAT Collaboration\footnote{\tt{http://www-glast.stanford.edu/cgi-bin/people}}}{
  address={Santa Cruz Institute for Particle Physics, University of California, Santa Cruz, CA 95064}
  ,altaddress={e-mail: pablo@scipp.ucsc.edu}
}

\begin{abstract}
The Large Area Telescope (LAT) on board the {\it Fermi} satellite has detected $\sim$120 pulsars above 100 MeV. While most $\gamma$-ray pulsars have spectra that are well modeled by a power law with an exponential cut-off at around a few GeV, some show significant pulsed high-energy (HE, $>$10 GeV) emission. I present a study of HE emission from LAT $\gamma$-ray pulsars and discuss prospects for the detection of pulsations at very high energies (VHE, $>$100 GeV) with ground-based instruments.
\end{abstract}

\maketitle


\section{Pulsars above 10 GeV: The EGRET view (1991-2000)}

Prior to the launch of {\it Fermi}, our best knowledge of the
high-energy (HE, $>$10 GeV) $\gamma$-ray sky came from EGRET. Although
{\it diffuse} emission accounted for the majority ($\sim$1300)
of the $\sim$1500 HE photons detected by EGRET in its 9
years in orbit, a few tens of such photons were coincident with five bright $\gamma$-ray pulsars: 10 from the
Crab (7 in the peaks), 4 from Vela (all in the peaks), 
10 from Geminga (5 in the peaks), 9 from B1706$-$44 (5 in the peaks), and 2 from B1951+32
(both in the peaks) \cite{Thompson05}. 

\section{{\slshape Fermi}-LAT catalogs: Past and Present}

Since its launch in 2008, the Large Area Telescope
(LAT\cite{Atwood09}) on {\it Fermi} has dramatically improved our knowledge of the $\gamma$-ray sky.  
The LAT has produced various catalogs in the last $\sim$3 years: the
Bright Source List (0FGL)~\cite{0FGL}, using 3 months of data to 
describe 205 ($>10\sigma$) $\gamma$-ray sources (30
pulsars). The First Pulsar Catalog (1PC)~\cite{1PC}, based on 6 months
of data, describing 46 $\gamma$-ray pulsars. The First/Second LAT
Source Catalogs (1FGL\cite{1FGL}, 2FGL\cite{2FGL}) using 11/24 months of data, and containing
1451/1873 sources (56/83 pulsars). Two catalogs, using 
36 months of data, are currently in preparation: The
{\it Fermi}-LAT Catalog of Sources above 10 GeV
(1FHL\cite{Paneque12}) describes the $\sim$500 ``Hard'' sources detected
by the LAT (25 coincident with pulsars) while the Second LAT Pulsar
Catalog (2PC) describes in depth the $\sim$120 LAT-detected ($>$100 MeV) $\gamma$-ray pulsars\cite{Celik12}.

\section{Search for HE emission from $\gamma$-ray pulsars}

We used 3-year data sets as in 1FHL and 2PC. We first tried to determine how many of the 25 sources
from 1FHL associated with LAT $\gamma$-ray
pulsars show {\it significant} pulsations (and can therefore be {\it
  identified} as pulsars). These 25 sources include: 5 EGRET
pulsars, 7 young (non-recycled) radio-selected $\gamma$-ray pulsars,
10 young (non-recycled) $\gamma$-selected pulsars, and 3 millisecond
$\gamma$-ray pulsars. 

Using the timing models from 2PC and {\tt gtsrcprob}, we
generated low energy (0.3--10 GeV) normalized\footnote{The area under
    the curve equals unity. We use 100 bins, so each bin
    width is 0.01 units of phase.}  weighted light curves ({\it
  templates}). HE histograms were obtained using unweighted
Front (Back) events within 0.6$^\circ$ (1.2$^\circ$) of the known
pulsar position, corresponding to $\sim$r95\% of the PSF. For each
pulsar, we defined, {\it a priori}, an ``off-pulse'' region, using
Bayesian Blocks~\cite{Scargle98} (See Figure 1, {\bf Left}), and evaluated the statistical
significance of the HE events coming from the ``pulsed'' region of the light curve. 

\begin{figure}[htbp]
\includegraphics[height=.225\textheight]{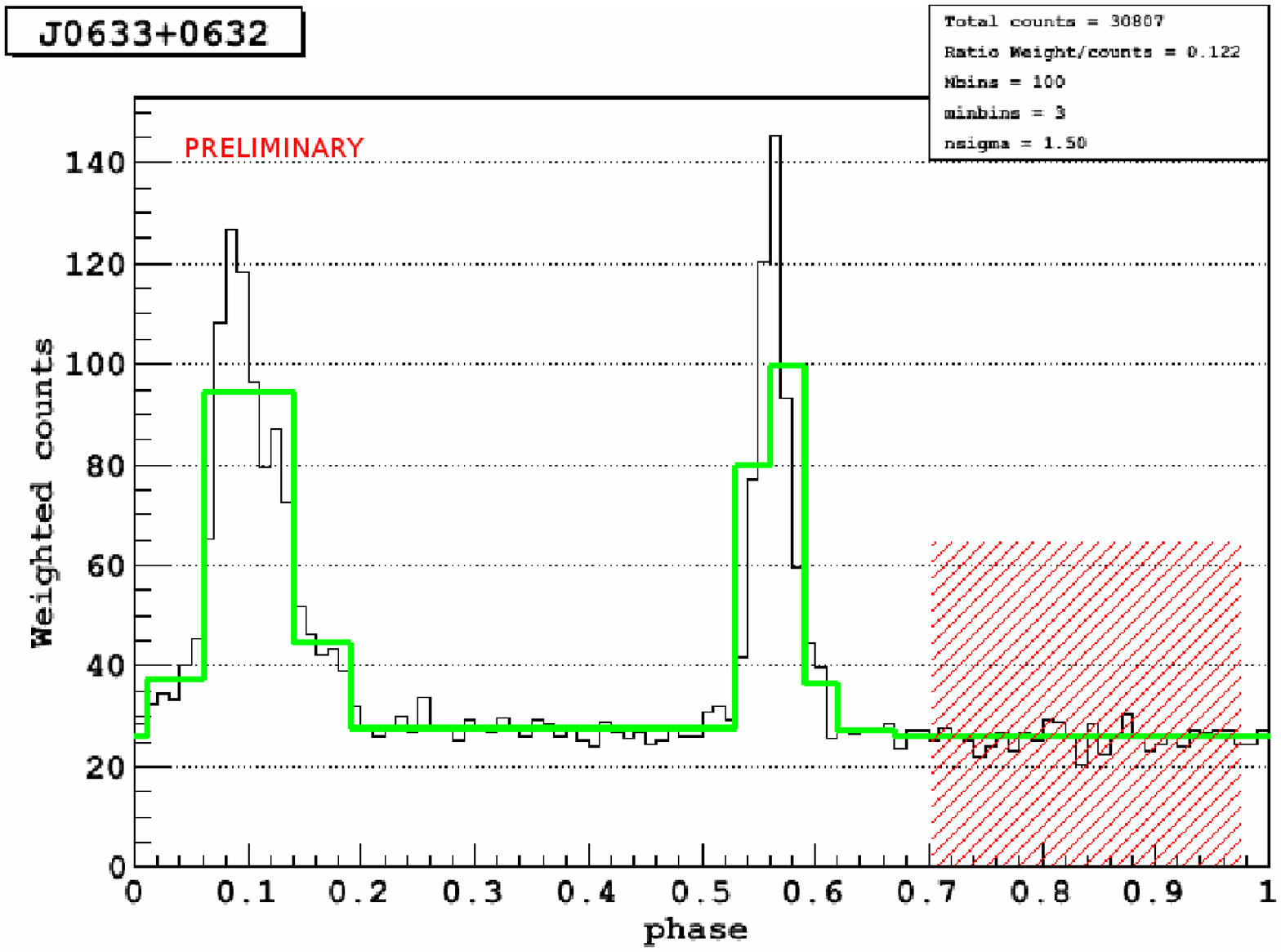}
\includegraphics[height=.25\textheight]{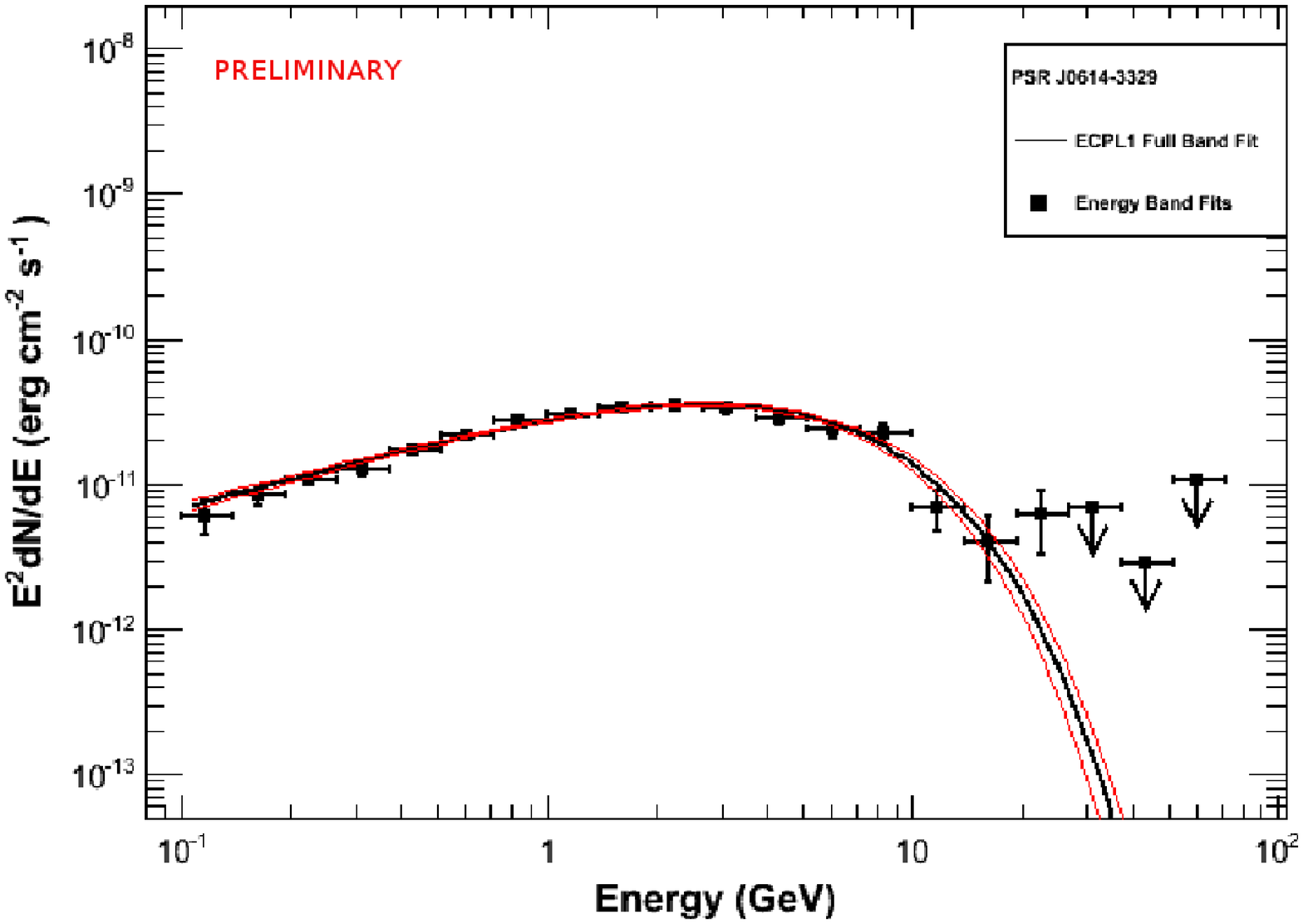}
\caption{{\bf Left} Preliminary off-pulse definition for
  PSR~J0633+0633 using Bayesian Blocks~\cite{Scargle98}. {\bf Right} Preliminary SED of PSR~J0614-3329, showing
  possible evidence for $>$10 GeV emission. }  
\label{fig6}
\end{figure}

We also selected a subset of the 117 $\gamma$-ray pulsars from
2PC which, based on their spectral energy distribution (SED, See
Figure 1, {\bf Right}), appear to emit above 10 GeV but did {\it not}
meet the criteria for inclusion in 1FHL. These spectrally-selected 2PC
pulsars not in the 1FHL catalog are: J0633+0632,
J1509--5850, J1747--2958, J1838--0537, J1954+2836, J2017+0603,
J2021+4026, J2238+5903, J2302+4442.

Our preliminary analysis shows that $\gtrsim$10 pulsars with
significant pulsed HE emission (including J0007+7303, Crab,  J0614-3329, Geminga,  Vela,
J1028-5819,  J1048-5832,  J1709-4429,  J1809-2332,  J2021+3651, and
J2032+4127). Several others require a more definitive analysis before
a firm detection can be claimed. Figure 2 shows the example of Geminga, where HE
pulsed emission is apparent, albeit with a very different pulse shape than what is seen at lower energies. Some of
the brightest pulsars (e.g. Geminga, Vela) show pulsed $>$25 GeV
emission (Figure 3), while at $>$50 GeV, the LAT starts running out of statistics, much
like EGRET did at $>$10 GeV (Figure 4). 

\begin{figure}
\includegraphics[height=.275\textheight]{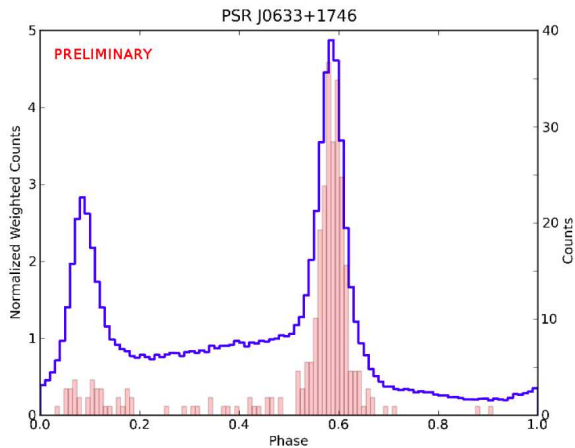}
\caption{Normalized weighted light curve of Geminga in the
  0.3--10 GeV energy range (blue, left scale) and unweighted HE light curve
  (pink, right scale). Weights based on 2PC spectral model.}
\end{figure}


\begin{figure}[htbp]
\includegraphics[height=.25\textheight]{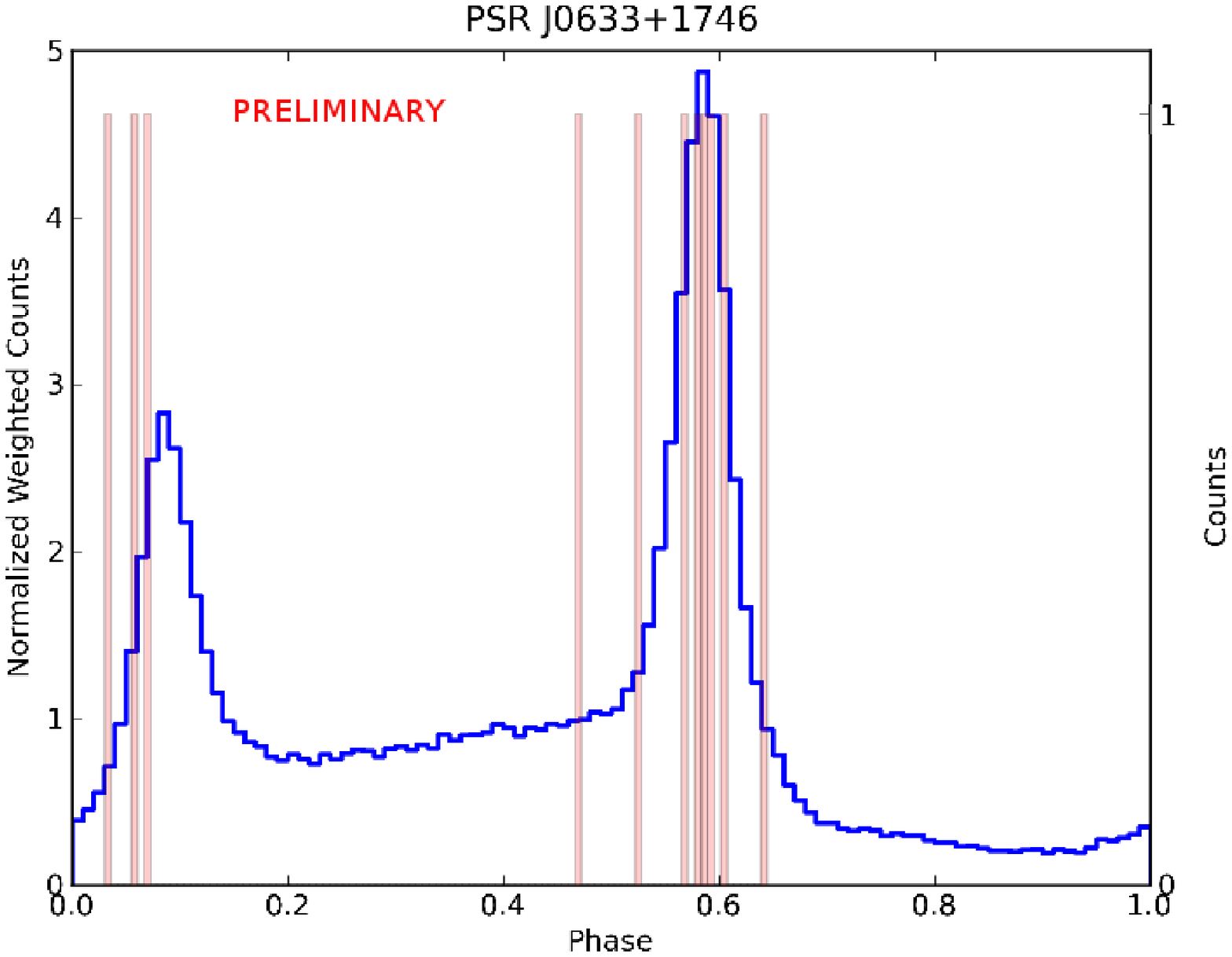}
\includegraphics[height=.25\textheight]{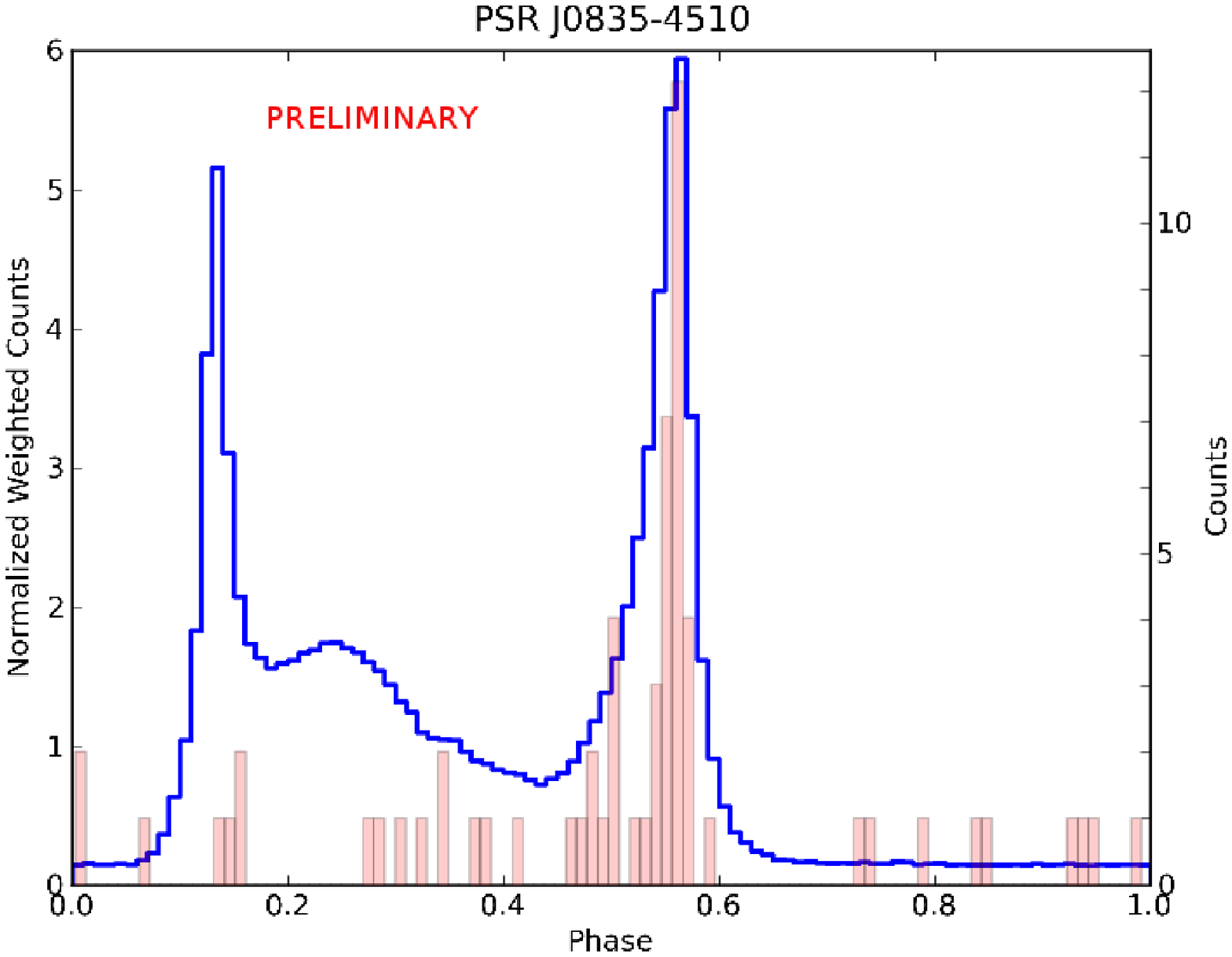}
\caption{Blue curves (left scale), same as Figure 2: normalized 0.3--10 GeV weighted light
  curve. Pink (unweighted) histogram (right scale) of $>$25 GeV events: {\bf Left} Geminga. {\bf Right} Vela.}
\label{fig6}
\end{figure}


\begin{figure}[htbp]
\includegraphics[height=.25\textheight]{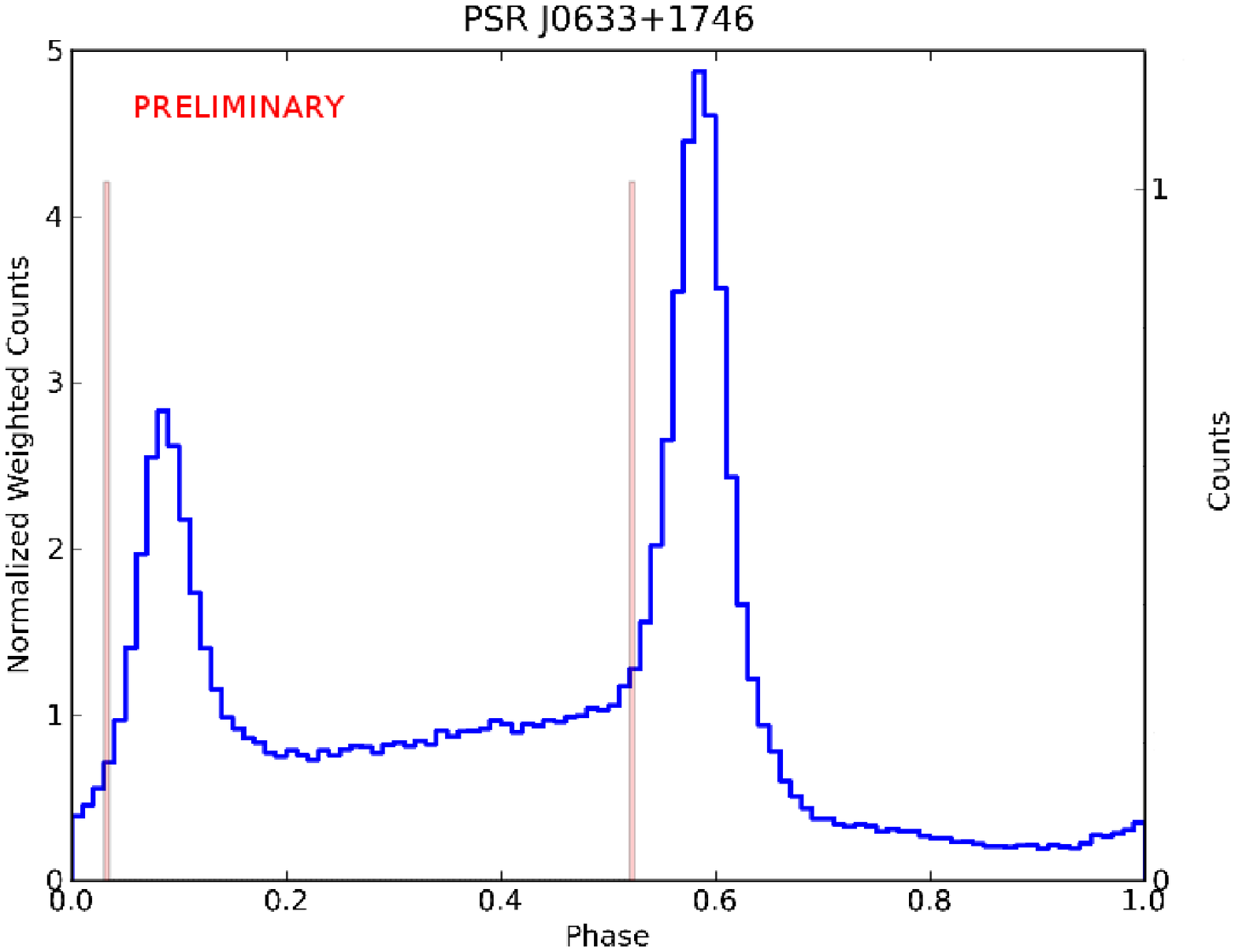}
\includegraphics[height=.25\textheight]{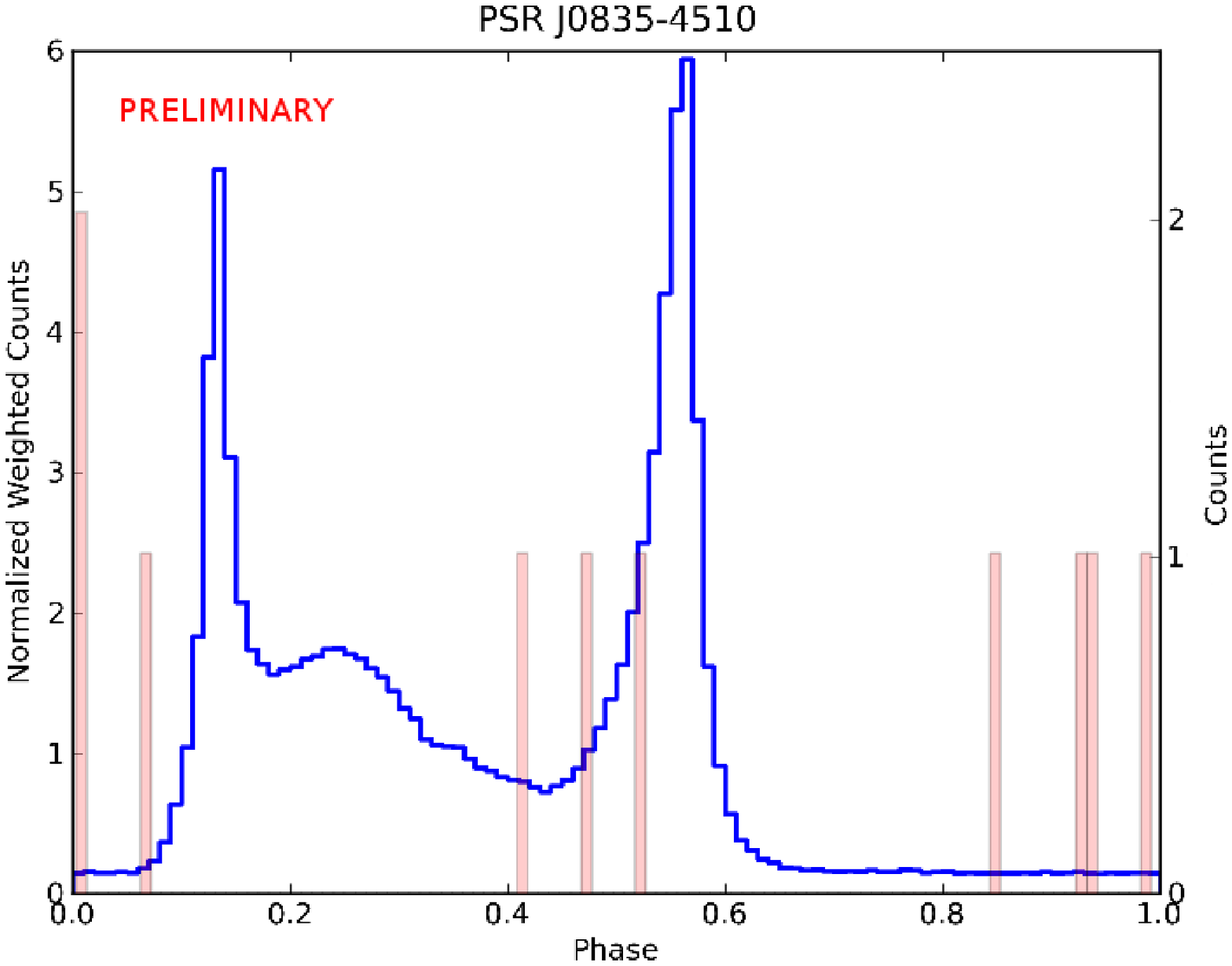}
\caption{Blue curves, same as Figures 2 and 3. Pink (unweighted) histograms (right
  scale) show $>$50 GeV events: {\bf Left} Geminga: 2 events above 50
  GeV (57 and 77 GeV). {\bf Right} Vela: 10 events above 50 GeV (5 events $>$100 GeV).}
\label{fig6}
\end{figure}



\section{Summary and Outlook}
The LAT has dramatically increased our knowledge of the hitherto
barely explored 10-100 GeV region of the $\gamma$-ray sky. A new LAT 
catalog in preparation (1FHL) will contain $\sim$500 HE sources, of which
25 are coincident with pulsars. In addition, the LAT has detected a
large number of $>$100 MeV $\gamma$-ray pulsars (to be described in the
upcoming 2PC), some of which show emission above 10 GeV. 
Future $\gamma$-ray pulsars may be discovered (e.g. in blind searches
or radio searches of LAT sources), but these will necessarily be
fainter than the brightest currently known. Top candidates for VHE
pulsations depend on many assumptions and spectral extrapolations from 10 GeV
upwards are notoriously unreliable. Thus, empirically speaking, the
bright EGRET pulsars (e.g. Geminga, Vela) remain among the best candidates for VHE
emission, while some of the newly-discovered bright LAT radio-quiet
$\gamma$-ray pulsars (e.g. CTA1) are also very promising. This preliminary
study used 3 years of data. More data are currently available and it
is important to keep in mind that in the statistics-limited
high-energy regime, LAT sensitivity improves faster (i.e. $\alpha$t)
than at lower energies, where backgrounds dominate. Future
improvements in reconstruction (e.g. Pass 8) could yield significant
increases in effective area at higher energies. Future TeV experiments
(e.g. CTA, HAWC) will complement and extend LAT observations in this
crucial energy window.


\begin{theacknowledgments}
The $Fermi$ LAT Collaboration acknowledges support from a number of agencies and institutes for both development and the operation of the LAT as well as scientific data analysis. These include NASA and DOE in the United States, CEA/Irfu and IN2P3/CNRS in France, ASI and INFN in Italy, MEXT, KEK, and JAXA in Japan, and the K.~A.~Wallenberg Foundation, the Swedish Research Council and the National Space Board in Sweden. Additional support from INAF in Italy and CNES in France for science analysis during the operations phase is also gratefully acknowledged.
\end{theacknowledgments}


\bibliographystyle{aipproc}   


\IfFileExists{\jobname.bbl}{}
 {\typeout{}
  \typeout{******************************************}
  \typeout{** Please run "bibtex \jobname" to optain}
  \typeout{** the bibliography and then re-run LaTeX}
  \typeout{** twice to fix the references!}
  \typeout{******************************************}
  \typeout{}
 }

\def \apjl {ApJL}
\def \apj {ApJ}
\def \apjs {ApJS}

\bibliography{ms}

\end{document}